\documentstyle [11pt]{article}
\def\fnote#1#2{\begingroup\def\thefootnote{#1}\footnote{#2}\addtocounter
{footnote}{-1}\endgroup}
\def\simlt{\stackrel{<}{{}_\sim}}
\def\simgt{\stackrel{>}{{}_\sim}}

\begin{document}

\hfill{IZTECH-P-14-02}

\vspace{36pt}

\begin{center}
{{\bf {Effects of Curvature-Higgs Coupling on Electroweak Fine-Tuning}}}

\vspace{36pt}
Durmu{\c s} Ali Demir\fnote{*}{Electronic address:
demir@physics.iztech.edu.tr}\\
{\em Department of Physics, {\.I}zmir Institute of Technology, IZTECH\\
TR35430, {\.I}zmir, Turkey}

\vspace{30pt}

\noindent
{\bf Abstract}
\end{center}

\noindent
It is shown that, nonminimal coupling between the Standard Model (SM) Higgs field 
and spacetime curvature, present already at the renormalizable level, 
can be fine-tuned to stabilize the electroweak scale against power-law ultraviolet 
divergences. The nonminimal coupling acts as an extrinsic stabilizer with no
effect on the loop structure of the SM, if gravity is classical. 
This novel fine-tuning scheme, which could also be interpreted within Sakharov's induced 
gravity approach, works neatly in extensions of the SM involving 
additional Higgs fields or singlet scalars. 

\vfill

\pagebreak

The discovery of a fundamental scalar \cite{higgs} by ATLAS and CMS experiments, and 
compatibility of this scalar with the SM Higgs boson \cite{higgs,ellis} prioritized the disastrous UV 
sensitivity of the Higgs boson mass \cite{scalar-1, scalar-2} as the foremost problem \cite{scalar-3} 
to be resolved. This is because, in the LHC searches  reaching out to energies fairly above 
the electroweak scale \cite{bsm}, the Higgs boson seems to lack any companion which would stabilize its mass. 
This means that the electroweak scale, set by the Higgs vacuum expectation value (VEV) 
\begin{eqnarray}
\label{Higgs-VEV}
v^2 = \frac{ - m_H^2}{\lambda_H}
\end{eqnarray}
that minimizes the Higgs potential 
\begin{eqnarray}
\label{higgs-pot}
V(H) = V_0 + m_H^2 H^{\dagger} H + \lambda_H \left(H^{\dagger}H\right)^2
\end{eqnarray}
for $m_H^2 < 0$ and $\lambda_H > 0$, is completely destabilized by the additive power-law 
quantum corrections $\delta m_H^2 \propto \Lambda_{UV}^2$ \cite{scalar-2}, where $\Lambda_{UV} \gg v$
is the UV scale which can be as high as $M_{Pl}$ if the SM is valid all the way up to the gravitational 
scale. 

The present paper will point out an exception to this inevitable destabilization by noting that the Higgs
field, being a doublet of fundamental scalars, necessarily develops the nonminimal 
Higgs-curvature interaction \cite{non-min}
\begin{eqnarray}
\label{nonmin-pot}
\Delta V(H,R) = \zeta R H^{\dagger} H
\end{eqnarray} 
with which the Higgs VEV in (\ref{Higgs-VEV}) changes to 
\begin{eqnarray}
\label{Higgs-VEV-zeta}
{v}^2 = \frac{- {m_H^2} - \frac{4 {\zeta} {V_0}}{{M_{Pl}^2}}}{{\lambda_H} + \frac{{\zeta} {m_H^2}}{{M_{Pl}^2}}}
\end{eqnarray}
and this new VEV can be stabilized by fine-tuning $\zeta$ to counterbalance the quadratic divergences
$\delta m_H^2 \propto \Lambda_{UV}^2$ with the quartic divergences $\delta V_0 \propto \Lambda_{UV}^4$. 
Quantum corrections to the SM parameters are independent of $\zeta$ if gravity is classical, and thus $\zeta$
acts as a gyroscope that stabilizes the electroweak scale against violent UV contributions. 
This novel fine-tuning scheme is in accord with Sakharov's induced gravity approach, and continues 
to hold also in extensions of the SM involving extra Higgs fields (additional Higgs doublets or singlet 
scalars or scalar multiplets belonging to larger gauge groups).

Below, we verify these observations by studying effects of the curvature-Higgs interaction (\ref{nonmin-pot}) on
the electroweak breaking, analyzing how fine-tuning of $\zeta$ leads to stabilization of
the electroweak scale, and determining implications of the mechanism for physics
beyond the SM.

In general, Higgs VEV is determined by the fields which can develop nontrivial backgrounds. Thus, only the Higgs VEV $v$
and the corresponding curvature scalar $R(v)$ matter at the tree level whilst all the fields coupling to the Higgs doublet count 
at the loop level. Explicating only its Higgs and curvature sections, the tree-level action is given by
\begin{eqnarray}
\label{action-0} 
S &\supset& \int d^{4}x \sqrt{-g} \Bigg\{\frac{1}{2} M_{Pl}^2 R - g^{\mu\nu} \left( {\mathcal{D}}_{\mu} H\right)^{\dagger}  \left({\mathcal{D}}_{\nu} H\right) - V(H)
- \Delta V(H,R)\nonumber\\ &-&\left[h_f \overline{F_L} H f_R + \mbox{H. C.}\right]\Bigg\}
\end{eqnarray}
where $F_L \sim SU(2)_L\otimes U(1)_Y$ and $f_R \sim U(1)_Y$ are quark and lepton fields, ${\mathcal{D}}_{\mu}$ is 
gauge-covariant derivative, and 
\begin{eqnarray}
\label{higgs-doublet}
H = \frac{1}{\sqrt{2}} \left(\begin{array}{l} \sqrt{2} \varphi^{+} \\ v + h + i \varphi^0\end{array}\right)
\end{eqnarray}
is the SM Higgs field encoding the Higgs boson $h$ and Goldstone bosons $\varphi^{+,-,0}$. Its VEV $v$
is determined to read as in (\ref{Higgs-VEV-zeta}) after a self-consistent solution of the curvature scalar
\begin{eqnarray}
\label{R-VEV-zeta}
R(v) = \frac{1}{M_{Pl}^2 - \zeta v^2}\left( 4 V_0 + 2 m_H^2 v^2 + \lambda_H v^4\right)
\end{eqnarray}
and the Higgs equation of motion
\begin{eqnarray}
\label{eom-higgs}
m_H^2 + \lambda_H v^2 + \zeta R (v) = 0
\end{eqnarray}
in the constant $v$ and $R(v)$ backgrounds. On physical grounds, $V_0 \neq 0$ as there exists no symmetry that dictates it.
Quite expectedly, the Higgs VEV in (\ref{Higgs-VEV-zeta}) tends to the usual Higgs VEV in (\ref{Higgs-VEV}) as $\zeta\rightarrow 0$. 
It is through the equations (\ref{R-VEV-zeta}) and (\ref{eom-higgs}) that the denominator in (\ref{Higgs-VEV-zeta}) reads 
${\lambda_H} + \left({{\zeta} {m_H^2}}\right)/{{M_{Pl}^2}}$ not just $\lambda_H$. Irrespective of if the Higgs field
couples minimally or nonminimally, the Higgs  VEV induces the Higgs boson mass $m_h^2 = 2 \lambda_H v^2$ 
properly if ${m_H^2} + \frac{ 4 {\zeta} {V_0}}{{M_{Pl}^2}} < 0$, and ensures strict masslessness of the Goldstone 
bosons on its equation of motion (\ref{eom-higgs}). (The spacetime curvature can influence 
symmetry breaking \cite{curvature}.)

In general, $\zeta$ is a free parameter that can be assigned appropriate values depending on 
the physical process under consideration. For instance, it is known to affect 
the LHC Higgs boson candidate \cite{non-min2} and weak boson scattering \cite{non-min3} 
for $\zeta \simeq 10^{15}$, and facilitate successful inflation for $\zeta \simeq 10^{4}$ \cite{higgs-inf}.

The Higgs VEV, which sets the electroweak scale and generates the masses of the SM particles, 
is the germinal physical observable. Its scale value is crucial for phenomenological success of 
the electroweak theory, and hence, its stability against quantum fluctuations is a vital issue by itself. Concerning 
the compution of quantum corrections to Higgs VEV, it is natural to construct the effective action \cite{eff-SM-1,visser} corresponding to
the tree-level action (\ref{action-0}) by incorporating into it the effects of the quantum fluctuations whose 
frequencies range from $\Lambda_{IR} \simgt v$ up to $\Lambda_{UV}$. Taking gravity classical to avoid nonrenormalizable 
quantum gravitational effects \cite{deser1}, the action (\ref{action-0}) is found to form a renormalizable setup if one sticks
to constant-curvature backgrounds \cite{curve} enabling direct comparison with the tree-level geometry in (\ref{R-VEV-zeta}). 
Then, one-loop quantum corrections to the parameters in Higgs potential (\ref{higgs-pot}) are given by 
\begin{eqnarray}
\label{V0-ren}
\delta{V_0} &=& \frac{1}{(4\pi)^2} \left[ \frac{1}{4} (n_F - n_B) \Lambda_{UV}^4 + 2 m_H^2 \Lambda_{UV}^2 + m_H^4 \log \frac{\Lambda_{IR}^2}{\Lambda_{UV}^2}\right]\\
\label{mH2-ren}
\delta{m_H^2} &=& \frac{3}{(4\pi)^2} \left[ \left(2 \lambda_H + \frac{g_Y^2}{4} + \frac{3 g_2^2}{4} - 2 h_t^2 \right)\Lambda_{UV}^2 + 2 \lambda_H m_H^2 \log \frac{\Lambda_{IR}^2}{\Lambda_{UV}^2}\right]\\
\label{lambdaH-ren}
\delta{\lambda_H} &=& \frac{3}{(4\pi)^2} \left[4 \lambda_H^2 + \frac{g_Y^4}{16} + \frac{g_2^2 g_Y^2}{8} + \frac{3 g_2^4}{16} + h_t^4 \right] \log \frac{\Lambda_{IR}^2}{\Lambda_{UV}^2}
\end{eqnarray}
where $h_t$ is top quark Yukawa coupling, $g_Y$ ($g_2$) is the hypercharge (isospin) gauge coupling, 
and $n_F$ ($n_B$) is the total number of fermions (bosons) in the SM. Unlike $\delta{V_0}$, $\delta{m_H^2}$ and $\delta \lambda_H$,
all of which are independent of $\zeta$, quantum corrections to gravity sector parameters
\begin{eqnarray}
\label{MPl2-ren}
\delta{M_{Pl}^2} &=& - \frac{1}{6 (4\pi)^2} \left[ \left(1 - 24 \zeta \right)\Lambda_{UV}^2
+ 4 m_H^2 \left(1 - 6 \zeta\right) \log \frac{\Lambda_{IR}^2}{\Lambda_{UV}^2}\right]\\
\label{zeta-ren}
\delta{\zeta} &=& - \frac{1}{(4\pi)^2} \left[ \lambda_H (1 - 6 \zeta) - \frac{3 g_Y^2}{8} - \frac{9 g_2^2}{8} + \frac{h_t^2}{12}\right] \log \frac{\Lambda_{IR}^2}{\Lambda_{UV}^2}
\end{eqnarray}
explicitly involve $\zeta$. Despite its quadratic divergence, $\delta {M_{Pl}^2}$ is a tiny quantum correction because $\Lambda_{UV} \simlt M_{Pl}$.
 
Pertaining to a renormalizable theory, the loop-level Higgs VEV is expected to have the same form as the tree-level VEV in (\ref{Higgs-VEV-zeta}). Thus, 
in response to the quantum corrections above, it changes by 
\begin{eqnarray}
\label{Higgs-VEV-shift} \delta{v^2} \simeq  \frac{3 Q(\zeta)}{(4\pi)^2 \lambda_H} \left( 2 h_t^2 - \frac{3}{4} g_2^2 - \frac{1}{4} g_Y^2 - 2 \lambda_H \right) \Lambda_{UV}^2
\end{eqnarray} 
as follows from (\ref{mH2-ren}) and (\ref{V0-ren}) after neglecting logarithmic UV contributions and dropping   
minuscule ${\mathcal{O}}\left(m_H^2/M_{Pl}^2\right)$ and ${\mathcal{O}}\left(V_0/M_{Pl}^4\right)$ terms. This correction differs from 
the well-known Veltman conditon \cite{scalar-2} by the loop factor
\begin{eqnarray}
\label{c-coef}
Q(\zeta) =   1 - \frac{\zeta (n_F - n_B)}{3 \left( 2 h_t^2 - \frac{3}{4} g_2^2 - \frac{1}{4} g_Y^2 - 2 \lambda_H \right)} \frac{\Lambda_{UV}^2}{M_{Pl}^2}
\end{eqnarray}
which is nothing but the ratio of the quartic divergence in ${V_0}$ to the quadratic divergence in ${m_H^2}$. It is this factor 
that differentiates between nonminimally- and minimally-coupled Higgs fields. Indeed, as $\zeta \rightarrow 0$, $Q(\zeta) \rightarrow 1$ and
Higgs VEV starts developing quadratic divergence, as expected from the familiar Higgs VEV (\ref{Higgs-VEV}) holding for minimally-coupled Higgs. 

A short glance at (\ref{Higgs-VEV-shift}) reveals that, as a new means not possible for minimally-coupled Higgs, one can suppress $\delta v^2$
to admissible level by imposing 
\begin{eqnarray}
\label{kosul}
|Q(\zeta)| \simlt \frac{ v^2 }{\Lambda_{UV}^2}
\end{eqnarray}
which amounts to an extreme fine-tuning of some 30 decimal places after comma. Thus, the nonminimal Higgs-curvature coupling 
$\zeta$ provides the SM with a novel fine-tuning mechanism for stabilizing the electroweak scale against power-law UV effects. 
This stabilization leads to stabilization of all particle masses, including that of the Higgs boson. 

Suppression of $Q(\zeta)$ in (\ref{kosul}) is accomplished by finely adjusting $\zeta$ in (\ref{c-coef}). Its duty is to counterbalance the 
quadratic divergence in ${m_H^2}$ with the quartic divergence in ${V_0}$. This is evident from the 
Veltman condition (\ref{Higgs-VEV-shift}) supplemented by (\ref{c-coef}). The workings of the fine-tuning in (\ref{kosul}) is best exemplified by 
the special value of $\zeta$
\begin{eqnarray}
\label{zeta-0}
\zeta_0 = \frac{1}{(n_F-n_B)} \left( 6 h_t^2 - 6 \lambda_H - \frac{3 g_Y^2}{4} - \frac{9 g_2^2}{4} \right) \frac{M_{Pl}^2}{\Lambda_{UV}^2}
\end{eqnarray}
for which $Q(\zeta_0)=0$. This specific nonminimal coupling has the numerical value $\zeta_0 \approx 1/15$ for $\Lambda_{UV} \approx M_{Pl}$.
It is smaller than the conformal value $1/6$ \cite{non-min} and much much smaller than the Higgs inflation value $10^{4}$ \cite{higgs-inf}. As a function of 
$\Lambda_{UV}$, $\zeta_0$ completely eradicates the power-law UV contribution (\ref{Higgs-VEV-shift}), and the concealed logarithmic corrections give 
the usual renormalization properties of the Higgs VEV. Obviously, smaller the $\Lambda_{UV}$ larger the $\zeta_0$ though there remains 
lesser and lesser need to fine-tuning if $\Lambda_{UV}$ gets closer and closer to the Fermi scale. 

The nonminimal coupling $\zeta$, as explicated in (\ref{zeta-ren}), receives additive logarithmic UV
corrections involving the SM gauge and Yukawa couplings. Hence, the tuned value of $\zeta$ in (\ref{kosul}),
as exemplified by (\ref{zeta-0}), receives small logarithmic corrections for which the value of $\zeta$
can be adjusted order by order in perturbation theory. The essential point is that it is the tree-level coupling $\zeta$,
not any of the SM parameters or momentum cutoffs $\Lambda_{UV/IR}$, which is finely tuned 
to achieve the suppression in (\ref{kosul}).

Quantum corrections to the SM parameters (Higgs, gauge and Yukawa sectors) do not involve $\zeta$. This is 
already evinced by (\ref{V0-ren}), (\ref{mH2-ren}) and (\ref{lambdaH-ren}). This observation is actually an
all-loop feature ensured by the classical nature of gravity, and makes certain that the SM maintains all of its IR/UV
quantum structures as if $\zeta$ does not exist. In other words, all the SM parameters run from scale to scale with 
no parameter tunings, coarse or fine. In essence, $\zeta$ behaves as an external gyroscope that maintains the bound (\ref{kosul})
despite various violent UV effects, and it is with this functionality that the matter and forces in 
the SM find themselves under optimal conditions for weak interactions to occur correctly.

To discuss further, we state that $\zeta$ fine-tuning can have a variety of implications 
for model building and phenomenology. Below we highlight some of them briefly:
\begin{itemize}
\item Our setup of cassical gravity plus quantized matter can be consistently interpreted 
within Sakharov's induced gravity \cite{sakharov,visser}. In this framework,  gravity is induced 
by matter loops as a long-distance effective theory, and this typically requires additional matter
multiplets to rightly induce the Planck scale $M_{Pl}$ \cite{sakharov,kazakh}. This means that, 
fine-tuning of $\zeta$ in (\ref{kosul}) might be deduced from symmetries of the non-SM 
matter multiplets. Interestingly, the non-SM matter here does not have to conform to 
supersymmetry or other UV-safe extensions of the SM \cite{wip}. 

\item The matter sector does not have to be precisely the SM. The fine-tuning mechanism 
here works also in extensions of the SM which include extra scalar fields provided
that each scalar assumes a nonminimal coupling to curvature as in (\ref{nonmin-pot}). The scalar
fields can be additional Higgs doublets, singlet scalars or multiplets of scalars
belonging to larger gauge groups. The VEV of each scalar is of the form in (\ref{Higgs-VEV-zeta}),
and can be fine-tuned individually without interfering with the VEVs of the remaining scalars.
These extended Higgs sectors can be probed at the LHC and other colliders \cite{lhc-higgs}.
The singlet scalars, in particular, can explain the cold Dark Matter \cite{dm} in Universe 
and enhance the invisible width of the Higgs boson \cite{singlet-higgs}.

\item The classical gravity assumption in the present work can be lifted to include quantum gravitational effects \cite{odintsov}. 
In this case, nonminimal coupling spreads into the SM parameters through graviton loops. Moreover,
this quantum gravitational setup is inherently nonrenormalizable \cite{deser1}. These factors can 
obscure the process of fine-tuning $\zeta$.

\item There have been various attempts \cite{hintli} to nullify the quadratic divergence in Higgs VEV by introducing singlet scalars.
This is now known to be not possible at all, even when vector-like fermions are included \cite{beste-canan}. Nevertheless,
nonminimal coupling between curvature scalar and some scalar fields can help stabilize both electroweak
and hidden scales as in (\ref{Higgs-VEV-shift}), and then masses of the particles in the SM and hidden 
sector get automatically stabilized. 

\item Throughout the discussions, cosmological constant problem \cite{zeldovich} is left aside as in
supersymmetry and other UV-safe extensions of the SM. The assumption is that it is a separate, 
independent naturalness problem pertaining to deep IR rather than electroweak or higher energy scales. 
The alleged mechanism that solves the cosmological constant problem must degravitate or dilute 
the vacuum energy at large distances. This can be accomplished presumably via modifications of gravity 
in the deep IR. In the present work, using (\ref{V0-ren}) in (\ref{R-VEV-zeta}) one finds 
$R(v)\sim {\mathcal{O}}\left(\Lambda_{UV}^{2}\right)$ which is some 120 orders of magnitude 
larger than its observational value of $R(exp) \simeq 10^{-47} {\rm eV}^2$ \cite{observe}, 
and modified gravitational dynamics becomes essential for diluting this curvature at large
distances. In this connection, one notes the empirical modifications of gravity 
which degravitate the vacuum enegry \cite{nima}  or canalize vacuum energy to 
gravitational constant instead of cosmological constant \cite{stress}. 
\end{itemize}

To conclude, we reiterate that the nonminimal curvature-Higgs coupling $\zeta$ plays a crucial role in stabilizing the electroweak scale.
If Higgs field were minimally-coupled, quadratic divergences in $m_H^2$ would induce the same divergences in 
$v^2$, simply because the latter is proportional to the former. Nevertheless, nonminimal curvature-Higgs interaction disrupts this 
proportionality by bringing $V_0$ into the game. Essentially, $\zeta$ causes Higgs
VEV to involve not only the Higgs mass parameter $m_H^2$ but also the vacuum energy $V_0$, 
and the quadratic divergence of the former can be counterbalanced with the quartic divergence of 
the latter if $\zeta$ is finely adjusted. Then, $\zeta$ acts as an external stabilizer 
that sets the electroweak scale without intervening with the quantum structure of the SM.

The various investigation directions commented above give an idea of how widespread the implications of 
the $\zeta$ fine-tuning scheme could be. It would be an important advancement to relate
the fine-tuning constraints on $\zeta$ at low-energies to the symmetries and spectra of 
the non-SM matter multiplets needed for inducing the Planck mass (matter multiplicity and $\Lambda_{UV}$ set $M_{Pl}$). 
On the other side, the LHC phenomenology of the extra Higgs fields and analysis of the singlet scalars
in regard to electroweak stability and Dark Matter phenomenology would be another important direction to 
explore. Last but not least, a fundamental understanding of the modified gravity models that 
render the vacuum weightless would be a crucial step towards completing the fine-tuning 
scheme presented in the present work. 

\bigskip

I thank ICTP Associate Program through which part of this work was carried out at the ICTP High Energy Section. I
thank to anonymous referee for interesting comments and suggestions.


\begin{thebibliography}{99}
\bibitem{curve}
The effective action contains higher-curvature terms $W^{\mu\nu\alpha\beta} W_{\mu\nu\alpha\beta}$ 
and $R^2$ (see \cite{eff-SM-1,visser}), which do not contribute to $R(v)$ in constant-curvature 
backgrounds. In other geometries, their contributions become negligible in small curvature limit.
\end{thebibliography}
\end{document}